\documentclass[aps,preprint,preprintnumbers,nofootinbib,showpacs]{revtex4}
\usepackage{graphicx}
\begin{document}
\title{Black hole, string ball, and $p$-brane production at hadronic
supercolliders}
\author{Kingman Cheung}
\email[Email:]{cheung@phys.cts.nthu.edu.tw}
\affiliation{National Center for Theoretical Sciences, National Tsing Hua
University, Hsinchu, Taiwan, R.O.C.}
\date{\today}

\begin{abstract}
In models of large extra dimensions, the string and Planck scales
become accessible at future colliders. When the energy scale is
above the string scale or Planck scale a number of interesting
phenomena occur, namely, production of stringy states, $p$-branes,
string balls, black hole, etc. In this work, we systematically
study the production cross sections of black holes, string balls,
and $p$-branes at hadronic supercolliders. We also discuss their
signatures. At the energy scale between the string scale $M_s$ and
$M_s/g_s^2$, where $g_s$ is the string coupling, the production is
dominated by string balls, while beyond
$M_s/g_s^2$ it is dominated by black holes. The production of a
$p$-brane is only comparable to black holes when the $p$-brane
wraps entirely on small extra dimensions. Rough estimates on the 
sensitivity reaches on the fundamental Planck scale $M_D$ are also obtained, 
based on the number of raw events.
\end{abstract}
\preprint{NSC-NCTS-020505}
\maketitle

\section{Introduction}

The standard model (SM) of particle physics, though it can fit most of the
present data, leaves a few fundamental problems unsolved, one of which
is the gauge hierarchy problem.
Since the second revolution of string theories, a crop of models with
extra dimensions have been proposed to solve various theoretical problems.
In an attractive
model of large extra dimensions or TeV quantum gravity (ADD model)
\cite{arkani},
the fundamental Planck scale can be as low as a few TeV.
This is made possible by localizing the SM particles on a
brane (using the idea of D-branes in Type I or II string theory),
while gravity is free to propagate in all dimensions.
The observed Planck scale ($\sim 10^{19}$ GeV) is then a derived quantity.
Extensive phenomenology studies have been carried out in recent years.
Signatures for the ADD model can be divided into two catergories:
sub-Planckian and trans-Planckian.  
The former is the one that was studied extensively, while
the latter just recently received more attention, especially black hole
production in hadronic collisions.  

A black hole (BH) has been illusive for decades, as we cannot directly
measure any properties of it, not to mention the production of black holes in
any terrestrial experiments.  This is due to the fact that in order to
produce black holes in collider experiments one needs a center-of-mass energy
above the Planck scale ($M_{\rm Pl}\sim 10^{19}$ GeV), 
which is obviously inaccessible at the moment.

In models of large extra dimensions, the properties of black holes
are modified and interesting signatures emerge
\cite{hole,banks,emp}. The fact that the fundamental Planck scale is
as low as TeV also opens up an interesting possibility of
producing a large number of black holes at collider experiments
(e.g. LHC) \cite{scott,greg}. Reference \cite{emp} showed that a
BH localized on a brane will radiate mainly in the brane, instead
of radiating into the Kaluza-Klein states of gravitons of the
bulk. In this case, the BH so produced will decay mainly into the
SM particles, which can then be detected in the detector.  This
opportunity has enabled investigation of the properties of BH at
terrestrial collider experiments.  There have been a number of
such studies
\cite{scott,greg,hoss,me,giddings,casa-lhc,hofmann,park,greg2,giudice,blei,solo,rizzo,RS-b} at
hadronic colliders.  A typical signature of the BH decay is a high
multiplicity, isothermal event, very much like a spherical
``fireball." On the other hand, BH production has also been studied
in cosmic ray experiments \cite{kay,feng,anch,ratt,ring,uhe,kowalski,han}. 
The ultrahigh energy cosmic rays
(UHECR) serve as a very energetic beam hitting on the atmosphere
as the target. The primary ingredient of UHECR is probably protons
or light nuclei, or could be photons or even neutrinos.  If it is
made up of protons, it could produce a large number of BH's up at
the top of the atmosphere, producing a giant air shower.  If it is
neutrinos, it could produce very interesting horizontal air
showers \cite{feng,anch,ratt,ring,uhe}, or black holes
within a km-sized neutrino telescope \cite{kowalski,han}.

An important quantity of a BH is its entropy $S_{\rm BH}$.  To fulfill the
thermodynamical description, a BH requires a large entropy of
order of 25 \cite{giddings}.  Such an entropy requirement implies
that the BH mass must be at least five times the fundamental
Planck scale \cite{giddings,me}.  
This mass requirement makes the BH production not
as large as previously calculated in a number of works
\cite{scott,greg,rizzo}, first pointed out in Ref. \cite{me}. In
addition, the signature of large multiplicity decay of a BH can only
happen when the entropy is large.  Even taking into account this
mass requirement, the event rate is still large enough for detection.  
On the other hand, there were arguments from Voloshin that the cross section
should be multiplied by an exponential factor 
$\sim \exp( - \frac{S_{\rm BH}}{n+1} )$ \cite{volo}
(we shall get back to this point later in Sec. III).  However,
this suppression factor becomes too severe for the production rate to be 
interesting, contrary to the conclusion of Ref. \cite{rizzo}, 
because of the large entropy requirement.
There have been continuous theoretical efforts to calculate the production 
and decay of BH's in particle collisions 
\cite{casa,kanti,kanti0,kim,ear,jevi,hsu,bilke,stoj}.

Other interesting trans-Planckian phenomena include string
balls \cite{emparan}, $p$-branes \cite{ahn,jain,feng-p}, 
and TeV string behaviors \cite{oda,tevstring}.
Dimopoulos and Emparan \cite{emparan} pointed out that when a BH reaches a
minimum mass, it transits into a state of highly excited and
jagged strings -- a string ball (SB). The transition point is at
\begin{equation}
M^{\rm min}_{\rm BH} = \frac{M_s}{g_s^2} \;,
\end{equation}
where $M_s$ is the string scale and $g_s$ is the string coupling.
Naively, SB's are stringy progenitors of BH's. The BH
correspondence principle states that properties of a BH with a
mass $M_{\rm BH}=M_s/g_s^2$ match those of a string ball 
with $M_{\rm SB}=M_s/g_s^2$. We can then equate the production
cross sections of SB and BH at the transition point.  In fact, we
shall use this argument to write down the cross section for the SB
at the transition point.   The existence of string balls could be
argued from the string point of view.  When the energy of the
scattering reaches the string scale, the scattering of particles is
no longer described by point-particle scattering but replaced by
string-string scattering.  As the energy goes further up, the
strings become highly excited, jagged and entangled string states,
and become like a string ball.   When the energy reaches the
transition point, it turns into a BH.  Previously, in the
discussion of a BH, we mentioned a large entropy requirement on the BH
in order for the object to be a BH.  Such a large mass requirement
makes the production cross section smaller than previously
thought.  Here in the case of SB's, the mass requirement is
substantially lower, thus the production rate is significantly
higher.  Hence, an SB is more interesting in the experimental point
of view if it decays with a distinct signature. Dimopoulos and
Emparan \cite{emparan} argued that the decay of a SB is similar to
that of a BH, i.e., a high multiplicity decay into the SM
particles, though in some intermediate stages the SB decays more
likely into bulk particles.

 Another
interesting trans-Planckian object is the $p$-brane.  A BH can be
considered a $0$-brane.  In particle collisions, if one considers
BH production, one should also consider $p$-brane production.  In
fact, the properties of $p$-branes reduce to those of a BH in the
limit $p\to 0$. In extra dimension models, in which there are
large extra dimensions and small extra dimensions of the size of the 
Planck length, let a $p$-brane wrap on $r$ small and $p-r$ large
dimensions. It was found \cite{ahn} that the production of
$p$-branes is comparable to BH's only when $r=p$, i.e., the
$p$-brane wraps entirely  on the small dimensions only.  If $r< p$, 
the production of $p$-branes would be suppressed by
powers of $(M_*/M_{\rm Pl})$, where $M_*$ is the fundamental scale of the
$4+n$ dimensions.  Therefore, here we only consider
the case in which $r=p$. The decay of $p$ branes is not well
understood. One interesting possibility is cascade into branes of
lower dimensions until they reach the dimension of zero.  Whether
the zero brane is stable depends on the model.  Another
possibility is the decay into brane and bulk particles, thus
experimentally the decay can be observed.  Or it can be a combination
of cascade into lower-dimensional branes and direct decays.

In this work,  we study the production rates of the BH's, SB's, and 
$p$-branes in hadronic collisions, with emphasis on the LHC and the VLHC.  
The organization is as
follows.  In the next section, we briefly describe the relation
between the fundamental Planck scale and the string scale.  In
Sec. III, we describe the production of BH's, SB's, and $p$-branes. In
Sec. IV, we show our numerical results for the LHC and VLHC.  We
discuss the decays in Sec. V and conclude in Sec. VI.

\section{Planck and string scales}

First let us address more clearly the configuration of the space-time.
Let there be $n$ total extra dimensions with $m$ small extra dimensions
and $n-m$ large extra dimensions.  When we say small extra dimensions, we
mean the size is of order of $1/M_*$, the fundamental Planck scale.
The observed 4D Planck scale $M_{\rm Pl}$ is then a derived quantity given by
\cite{arkani}
\begin{equation}
\label{nm}
M_{\rm Pl}^2 = M_*^{2+n} \, V_{m} \, V_{n-m} \;,
\end{equation}
where $V_{m}$ and $V_{n-m}$ are the volumes of the extra $m$ and $n-m$ 
dimensions, respectively, given by
\begin{equation}
V_m = L_m^m \equiv \left( \frac{ l_m}{M_*}\right)^m\,; \qquad
V_{n-m} = L_{n-m}^{n-m} \equiv \left( \frac{ l_{n-m}}{M_*}\right)^{n-m}\,,
\end{equation}
where we have expressed the lengths $L_m, L_{n-m}$ in units of Planckian length
$1/M_*$.
\footnote{In the case of toroidal compactification, the length
$L_{i} =2\pi R_{i}\; (i=m,n-m)$ where $R_{i}$ is the radius of the torus.}
Suppose the small extra dimension has the size of $L_m \sim 1/M_*$, i.e.,
$l_m \sim 1$,  then
\begin{equation}
\label{n}
M_{\rm Pl}^2 = M_*^{2} \; \left( l_{n-m}  \right)^{n-m} \;.
\end{equation}
The fundamental Planck scale $M_*$ is lowered to the TeV range if the size
$L_{n-m}$ is taken to be very large, of order $O({\rm mm})$.

The relation of the observed Planck scale to the string scale $M_s$ is given
by \cite{tye}
\begin{equation}\label{nmm}
M_{\rm Pl}^2 \sim \frac{M_s^{2+n}}{g_s^2 } \,V_{m} \,V_{n-m} \;,
\end{equation}
where we again take the small extra dimensions of size $L_m \sim 1/M_*$.
{}From Eqs. (\ref{nm}) and (\ref{nmm}) we can relate the string scale with
the fundamental Planck scale as
\begin{equation} \label{ms-m*}
M_*^{2+n} \sim  \frac{M_s^{2+n}}{g_s^2 } \;,
\end{equation}
where we take the proportional constant of order $O(1)$, which depends on
different compactification configurations.  We shall also use the more
conventional definition of the fundamental 
Planck scale $M_D$ related to $M_*$ by
\begin{equation} \label{m*-md}
M_D^{n+2} = \frac{( 2\pi )^{n} }{ 8 \pi G_{4+n}} =
              \frac{( 2\pi )^{n} }{ 8 \pi} \, M^{n+2}_* \;,
\end{equation}
where $G_{4+n}$ is the gravitational constant in $D=4+n$ dimensions
(used in the Einstein equation:
${\cal R}_{AB} - \frac{1}{2} g_{AB} {\cal R} = -8\pi G_{4+n} T_{AB}$).
Then $M_D$ is related to the string scale $M_s$ as
\begin{equation} \label{ms-md}
M_D^{n+2} = K \,\frac{M_s^{n+2}}{g_s^2 } \;,
\end{equation}
for a constant $K$ of order $O(1)$.  In the next section we shall simply
use $K=1$ for discussions and cross-section calculations.

\section{Production}

\subsection{Black holes}

A black hole is characterized by its Schwarzschild radius $R_{\rm BH}$, which
depends on the mass $M_{\rm BH}$ of the BH.  A simplified picture for
BH production is as
follows.  When the colliding partons have a center-of-mass (c.o.m.) energy
above some thresholds of order of the
Planck mass and the {\it impact parameter} less
than the Schwarzschild radius $R_{\rm BH}$, a BH is formed
and almost at rest in the c.o.m. frame.  The BH so produced will decay
thermally (regardless of the incoming particles) and thus isotropically
in that frame.

This possibility was first investigated for the LHC in 
Refs. \cite{scott,greg}. In Refs. \cite{scott,greg,hoss}, black
hole production in hadronic collisions is calculated in $2\to 1$
subprocesses: $ij\to\; {\rm BH}$, where $i,j$ are incoming
partons.  The black hole so produced is either at rest or
traveling along the beam pipe such that its decay products (of
high multiplicity) have a zero net transverse momentum ($p_T$).
Giddings and Thomas \cite{scott} and Dimopoulos and Landsberg
\cite{greg} demonstrated that a BH so produced will decay with a
high multiplicity.

In Ref. \cite{me}, we pointed out the ``$ij\to {\rm BH}+\;{\rm others}$''
 subprocesses,
such that the BH is produced with a large $p_T$ before it decays.
The ``$ij\to {\rm BH}+\;{\rm others}$'' subprocesses
can be formed when the c.o.m. energy of the colliding particles
is larger than the BH mass; the
excess energy will be radiated as other SM particles.
\footnote{
Another viewpoint on BH formation is as follows.  The BH is formed
when the two partons approach each other in a very short distance
($ < R_{\rm BH}$), and everything, including the ``extra partons,'' will
be hidden behind the horizon.  
Thus, the entire energy is contained within the BH,
which is considered a quantum state rather than a particle.  In this
picture, there is no transverse momentum given to the BH, unless by
the initial-state radiation of the incoming partons. This was already
studied by a Monte Carlo approach in Ref. \cite{greg}.
}
In such subprocesses, the ``others'' are just the ordinary SM particles and
usually of much lower multiplicity than the decay products of the BH.
Therefore, the signature is as follows: on one side of the event
there are particles of high multiplicity (from the decay of the BH), the
total $p_T$ of which is balanced by a much lower number of particles on
the other side.  Such a signature is very clean and
should have very few backgrounds.

The next natural question to ask is how large the event rate is.
Collider phenomenology is only possible if the event rate is large
enough, especially if we want to study BH properties. It was
first pointed out in Ref. \cite{greg} that the BH production rate
is so enormous that the LHC is in fact a BH factory.  However, this
has not taken into account the entropy factor of the BH.  It was
shown in Ref. \cite{me} that with a large entropy requirement
($S_{\rm BH} \agt 25$) the BH production rate decreases
substantially, but still affords enough events for detection.  For
example, the production cross section is as high as 1 pb for
$M_D=1.5$ TeV and $n=4$ with $M_{\rm BH}^{\rm min}=5 M_D$ at the LHC.  
This implies $10^5$ events with an integrated luminosity of 100
fb$^{-1}$. Of course, if we relaxed this entropy constraint, the
production cross sections would be increased tremendously, but the
cross sections have to be interpreted with care because of the
presence of large string effects in this regime.

The Schwarzschild radius $R_{\rm BH}$ of a BH of mass $M_{\rm BH}$ in
$4+n$ dimensions is given by \cite{myers}
\begin{equation}
\label{r}
R_{\rm BH} = \frac{1}{M_D}\; \left (
\frac{M_{\rm BH}}{M_D} \right)^{\frac{1}{n+1}}\;
\left( \frac{ 2^n \pi^{ \frac{n-3}{2}} \Gamma(\frac{n+3}{2} )}{n+2}
\right )^{\frac{1}{n+1}}
= \frac{1}{M_D}\; \left (
\frac{M_{\rm BH}}{M_D} \right)^{\frac{1}{n+1}}\; f(n)
\;,
\end{equation}
where $f(n)$ is introduced for convenience and
 $M_D$ is the fundamental Planck scale in the model of large extra
dimensions already defined in Eq. (\ref{m*-md}).
The radius $R_{\rm BH}$ is much smaller than the size of the extra dimensions.
BH production is expected when the colliding partons with a center-of-mass
energy $\sqrt{\hat s} \agt M_{\rm BH}$ pass within a distance less than
$R_{\rm BH}$.  A black hole of mass $M_{\rm BH}$ is formed and the rest of
the energy, if there is any, is radiated as ordinary SM particles.
This semiclassical argument calls for a geometric approximation 
for the cross section for producing a BH of mass $M_{\rm BH}$ as
\begin{equation}
\label{geo}
\sigma(M_{\rm BH}^2 ) \approx \pi R_{\rm BH}^2 \;.
\end{equation}
In the $2\to 1$ subprocess, the c.o.m. energy of the colliding partons is just
the same as the mass of the BH, i.e., $\sqrt{\hat s}=M_{\rm BH}$, which
implies a subprocess cross section
\begin{equation}
\label{2to1}
\hat \sigma(\hat s) = \int \; d\left(\frac{M_{\rm BH}^2}{\hat s}\right )\;
\pi R_{\rm BH}^2 \; \delta\left( 1 - M^2_{\rm BH}/\hat s \right )
= \pi R_{\rm BH}^2 \;.
\end{equation}
On the other hand, for the $2\to k (k\ge 2)$ subprocesses the subprocess
cross section is
\begin{equation}
\label{2to2}
\hat \sigma(\hat s) = \int^{1}_{ (M^2_{\rm BH})_{\rm min}/\hat s} \;
d\left(\frac{M_{\rm BH}^2}{\hat s}\right )\; \pi R_{\rm BH}^2 \;.
\end{equation}

\begin{figure}[th!]
\includegraphics[width=5in]{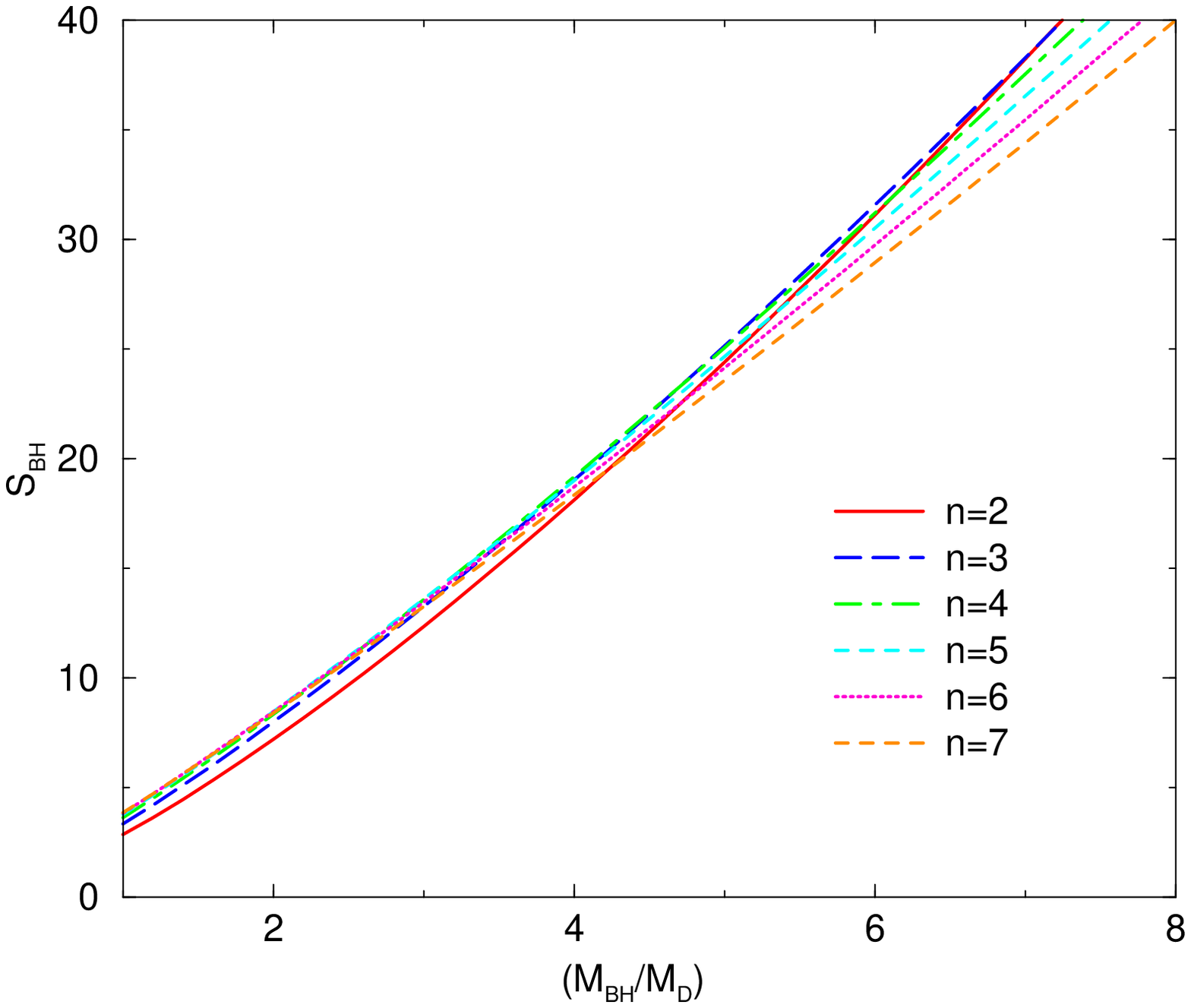}
\caption{\small
\label{sbh}
The entropy $S_{\rm BH}$ of a black hole vs the ratio $(M_{\rm BH}/M_D)$
in $4+n$ dimensions.}
\end{figure}

Another important quantity that characterizes a BH is its entropy given by
\cite{myers}
\begin{equation}
\label{entropy}
S_{\rm BH} = \frac{4\pi}{n+2}\; \left (
\frac{M_{\rm BH}}{M_D} \right)^{\frac{n+2}{n+1}}\;
\left( \frac{ 2^n \pi^{ \frac{n-3}{2}} \Gamma(\frac{n+3}{2} )}{n+2}
\right )^{\frac{1}{n+1}} \;.
\end{equation}
The variation of $S_{\rm BH}$ versus the ratio $M_{\rm BH}/M_D$ is
shown in Fig. \ref{sbh}.
 To ensure the validity of the above
classical description of a BH \cite{giddings}, the entropy must be
sufficiently large, of order 25 or so. From the figure we can see
that when $M_{\rm BH}/M_D \agt 5$, the entropy $S_{\rm BH}\agt
25$.  Therefore, to avoid getting into the nonperturbative regime
of the BH and to ensure the validity of the semiclassical
formula, we restrict the mass of the BH to be $M_{\rm BH} \ge y M_D$,
where $y \equiv M_{\rm BH}^{\rm min}/M_D$ is of order 5.

Voloshin \cite{volo}
pointed out that the semiclassical argument for the BH
production cross section is not given by the geometrical cross-section
area, but, instead,  suppressed by an exponential factor,
\begin{equation}
\label{supp}
\exp \left( - \frac{S_{\rm BH}}{n+1} \right ) \;.
\end{equation}
There are, however, counter arguments \cite{giddings,emparan} that
the simple geometric formula should be valid.
\footnote{
The controversy of using the exponential suppression factor seems
to be resolved by now.  The naive geometric cross section is correct,
as pointed out or even derived by various authors in Refs. 
\cite{casa,kanti,kanti0,kim,ear,jevi,hsu,bilke,emparan}.  
In particular, Ref. \cite{ear} explicitly pointed out a logical error by
Voloshin and derived the geometric cross-section formula.}
In Ref. \cite{me}, we have considered both forms of cross sections:
the naive $\pi R_{\rm BH}^2$ and the $\pi R_{\rm BH}^2$ multiplied
with the exponential factor of Eq. (\ref{supp}).
The suppression factor renders the cross section too small for detection,
because the exponential factor contains the entropy $S_{\rm BH}$,
which has to be sufficient large (e.g., $\agt 25$) to define a black hole.
The suppression is more than two orders of magnitude and, therefore,
we shall not be concerned with this suppression factor anymore.
This seems to contradict the results of Ref. \cite{rizzo},
in which Rizzo concluded that even when Voloshin's suppression
factor is included, the cross section is still large enough,
because he did not impose a large entropy requirement on the
validity of the BH. If he had done so, he would also have gotten a very
large suppression. Nevertheless, careful interpretation is needed
if the BH has only a small entropy.

\subsection{String balls}

Dimopoulos and Emparan \cite{emparan} pointed out that when a BH reaches
a minimum mass,  it transits into a state of highly excited
and jagged strings, dubbed string ball. They are the stringy progenitors
of BH's and share some properties of BH's, such as large production cross
sections at hadronic supercolliders and similar signatures when they decay.
They made an important observation \cite{emparan} that the minimum mass
$M_{\rm BH}^{\rm min}$ (the transition point) above which a BH can be treated
general-relativistically 
is $M_s/g_s^2$, where $M_s$ and $g_s$ are the string
scale and the string coupling, respectively.  Below this transition point,
the configuration is dominated by string balls.  Since the mass of
a string ball is lower than a BH, the corresponding production cross
section is larger than that of a BH.  Thus, at the LHC, string ball production
may be more important.

According to the BH correspondence principle, the properties of a BH with
a mass $M_{\rm BH}=M_s/g_s^2$ match those of a string ball 
with a mass $M_{\rm SB}=M_s/g_s^2$.  Therefore, the production cross
section of a string ball or a BH should be smoothly joined at $M_{\rm BH}=
M_s/g_s^2$, i.e.,
\[
\left. \sigma(SB) \right|_{M_{SB} = M_s/g_s^2} =
\left. \sigma(BH) \right|_{M_{BH} = M_s/g_s^2} \;.
\]
The production cross section for string balls with mass 
between the string scale $M_s$
and $M_s/g_s$ grows with $s$ until $M_s/g_s$, beyond which, due to unitarity,
it should stay constant.  Therefore, we can use the BH cross section and match
to the string ball cross section at the transition point $M_s/g_s^2$.
This string ball cross section then stays constant between $M_s/g_s$ and
$M_s/g_s^2$.  Then below $M_s/g_s$ the string ball cross section grows like
$M_{\rm SB}^2/M_s^4$.

\begin{figure}[th!]
\includegraphics[width=5in]{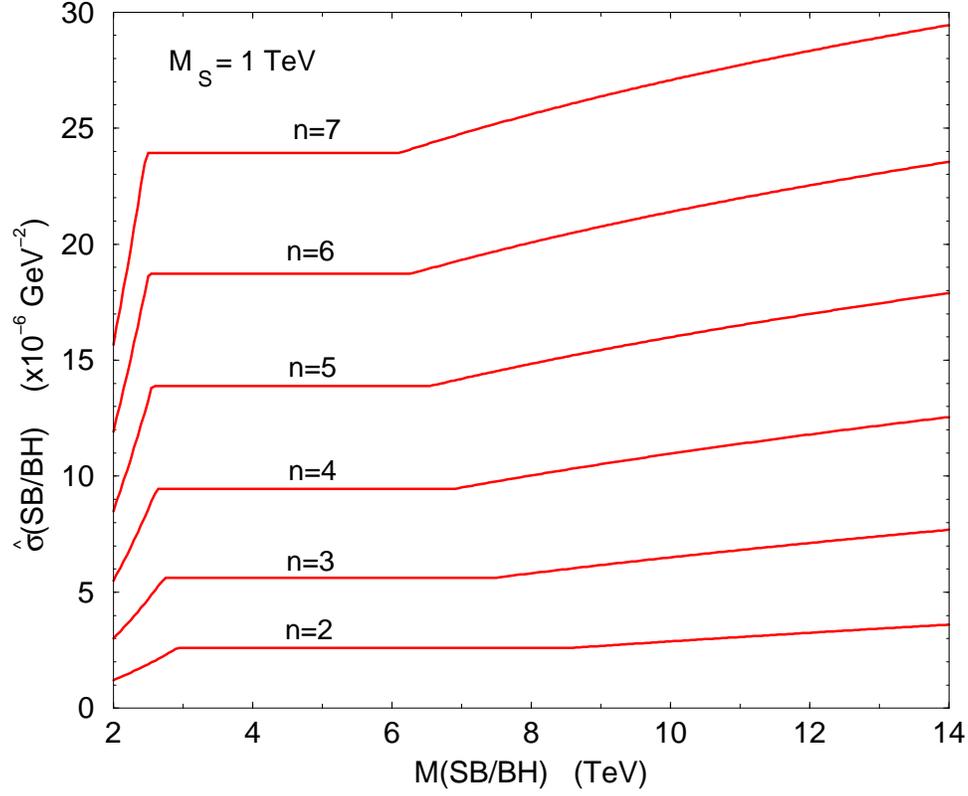}
\caption{\small
\label{sigma-hat}
The subprocess cross section $\hat \sigma$(SB/BH) for string ball or black
hole vs the mass of SB or BH.  Here we have used a string scale $M_s=1$
TeV, and we require the SB-BH correspondence point at $M_s/g_s^2 = 5 M_D$,
where $M_D$ is related to $M_s$ by $M_s = M_D g_s^{2/(n+2)}$.
}
\end{figure}

The cross sections for the SB or BH are given by
\begin{equation}
\hat\sigma ({\rm SB/BH}) = \left \{ \begin{array}{ll}
\frac{\pi}{M_D^2}\; \left (
\frac{M_{\rm BH}}{M_D} \right)^{\frac{2}{n+1}}  \left[ f(n) \right ]^2 &
                         \frac{M_s}{g_s^2} \le M_{\rm BH} \\
\frac{\pi}{M_D^2}\; \left (
\frac{M_s/g_s^2}{M_D} \right)^{\frac{2}{n+1}}  \left[ f(n) \right ]^2
 = \frac{ \pi}{M_s^2}  \left[ f(n) \right ]^2 &
                \frac{M_s}{g_s} \le M_{\rm SB} \le \frac{M_s}{g_s^2} \\
\frac{ \pi g_s^2 M_{\rm SB}^2 }{M_s^4}  \left[ f(n) \right ]^2 &
                M_s \ll M_{\rm SB} \le \frac{M_s}{g_s}
\end{array}
 \right. \;,
\end{equation}
in which we have set $M_D^{n+2} = \frac{M_s^{n+2}}{g_s^2 }$ as in
Eq. (\ref{ms-md}) with $K=1$. A graphical presentation of these
cross sections is shown in Fig. \ref{sigma-hat} for $n=2-7$.

In the next section, when we calculate the production cross sections for
BH's and SB's, we use the above equation, together with Eq. (\ref{ms-md})
 with $K=1$.  The
production then depends on the following parameters: $M_s, g_s, n$, and $M_D$.
The $M_D$ can be determined by Eq. (\ref{ms-md}).  We also require that at
the BH-SB transition point, $M_{\rm BH}^{\rm min}=M_s/g_s^2$, 
the mass of the BH is already at $5 M_D$
(this ensures that the BH has a sufficiently large entropy $\sim 25$
\cite{giddings}).  Therefore, the production cross sections depends on
$M_s$ and $n$ only.
We shall present the results in terms of $M_D$ and $n$ for easy comparison
with existing literature.

\subsection{$p$-Branes}

A black hole can be considered a zero-brane.  In principle, higher-dimensional
objects, e.g., $p$-branes ($p$B),
can also be formed in particle collisions,  in particular
when there exist small extra dimensions of the size $\sim 1/M_*$ in addition
to the large ones of the size $\gg 1/M_*$.  It was pointed out
by Ahn {\it et al.} \cite{ahn} that the production cross section
of a $p$-brane completely wrapped on the small extra dimensions is larger than
that of a spherically symmetric black hole. A similar situation is true
in cosmic ray experiments \cite{jain,feng-p}.

Consider an uncharged and static $p$-brane with a mass $M_{p \rm B}$ in 
$(4+n)$-dimensional 
space-time ($m$ small Planckian size and $n-m$ large size extra
dimensions such that $n \ge p$).
Suppose the $p$-brane wraps on $r (\le m)$ small extra dimensions and on
$p-r (\le n-m)$ large extra dimensions.
Then the ``radius'' of the $p$-brane is
\begin{equation}
\label{rp}
R_{p \rm B} = \frac{1}{\sqrt{\pi} M_*} \, \gamma(n,p) \,
V_{p\rm B}^{ \frac{-1}{1+n-p} } \,
\left( \frac{M_{p\rm B}}{M_*} \right)^{ \frac{1}{1+n-p} } \;,
\end{equation}
where $V_{p\rm B}$ is the volume wrapped by the $p$-brane in units of the
Planckian length.  
Recall from Eq. (\ref{nm}), $M_{\rm Pl}^2 = M_*^2 l_{n-m}^{n-m} l_m^m$,
where $l_{n-m}\equiv L_{n-m}\,M_{*}$ and $l_m \equiv L_m\,M_{*}$ 
are the lengths of the size of the large and small
extra dimensions in units of Planckian length ($\sim 1/M_*$).  Then
$V_{p\rm B}$ is given by
\begin{equation}\label{vp}
V_{p \rm B} = l_{n-m}^{p-r} \, l_m^{r} \approx \left( \frac{M_{\rm Pl}}{M_*}
\right )^{ \frac{2(p-r)}{n-m} } \;,
\end{equation}
where we have taken $l_m \equiv L_m \,M_{*} \sim 1$.  
The function $\gamma(n,p)$ is given by
\begin{equation}\label{gamma}
\gamma(n,p) = \left[ 8 \Gamma \left( \frac{3+n-p}{2} \right) \sqrt{
\frac{1+p}{(n+2)(2+n-p)} } \right ]^{\frac{1}{1+n-p} } \;.
\end{equation}
The $R_{p\rm B}$ reduces to the $R_{\rm BH}$ in the limit $p=0$.

The production cross section of a $p$-brane is similar to that of BH's, based
on a naive geometric argument \cite{ahn}.  When the partons collide with a
center-of-mass energy $\sqrt{\hat s}$ larger than the fundamental Planck
scale and an impact parameter less than the size of the $p$-brane, a $p$-brane
of mass $M_{p\rm B} \le \sqrt{\hat s}$ can be formed.  That is, 
\begin{equation}
\hat \sigma (M_{p\rm B}) = \pi R^2_{p\rm B} \;.
\end{equation}
Therefore, the production cross section for a $p$-brane is the same as BH's
in the limit $p=0$ (i.e., a BH can be considered a $0$-brane).
In $2\to1$ and $2\to k \;(k\ge 2)$ processes, the parton-level cross sections
are given by similar expressions in Eqs. (\ref{2to1}) and (\ref{2to2}),
respectively.

In Eq. (\ref{rp}), we can see that the radius of a $p$-brane is suppressed by
some powers of the volume $V_{p\rm B}$ wrapped by the $p$-brane.  It is then
obvious that the production cross section is largest when $V_{p\rm B}$ is
minimal, in other words, the $p$-brane wraps entirely on the small extra
dimensions only, i.e., $r=p$.  When $r=p$, $V_{p\rm B}=1$.
We can also compare the production cross section of $p$-branes with BH's.
Assuming that their masses are the same and the production threshold
$M^{\rm min}$ is the same, the ratio of cross sections is
\begin{equation}
\label{R}
R \equiv
\frac{\hat \sigma (M_{p\rm B}=M)}{\hat \sigma (M_{\rm BH}=M)}
= \left ( \frac{M_*}{M_{\rm Pl}} \right)^{\frac{4(p-r)}{(n-m)(1+n-p)}}
\, \left(\frac{M}{M_*} \right )^{ \frac{2p}{(1+n)(1+n-p)} } \,
\left( \frac{\gamma(n,p)}{\gamma(n,0)} \right )^2 \;.
\end{equation}
In the above equation, the most severe suppression factor is in the first
set of parentheses on the right-hand side.  Since we are considering physics of
TeV $M_*$, the factor $(M_*/M_{\rm Pl}) \sim 10^{-16}-10^{-15}$.  
Thus, the only
meaningful production of a 
$p$-brane occurs for $r=p$, and then their production
is comparable.  In Table \ref{pbrane}, we show this ratio for various
values of $n$ and $p$.

\section{Production at the LHC and VLHC}

The production of BH's and SB's depends on $M_s,n,M_D,g_s$, but they
are related by Eq. (\ref{ms-md}). Since we also require the
transition point ($M_s/g_s^2$) at $5M_D$, we can therefore solve
for $M_s$ and $g_s$ for a given pair of $M_D$ and $n$.  
We present the
results in terms of $M_D$ and $n$.  The minimum mass requirement
for the SB is set at $2M_s$.  The production of a $p$-brane also
depends on $m$ and $r$.  For an interesting level of event rates,
$r$ has to be equal to $p$, i.e., the $p$-brane wraps entirely on
small (of Planck length) extra dimensions.  So after setting all
parameters, we are ready to present our numerical results.

\begin{figure}[th!]
\includegraphics[width=5in]{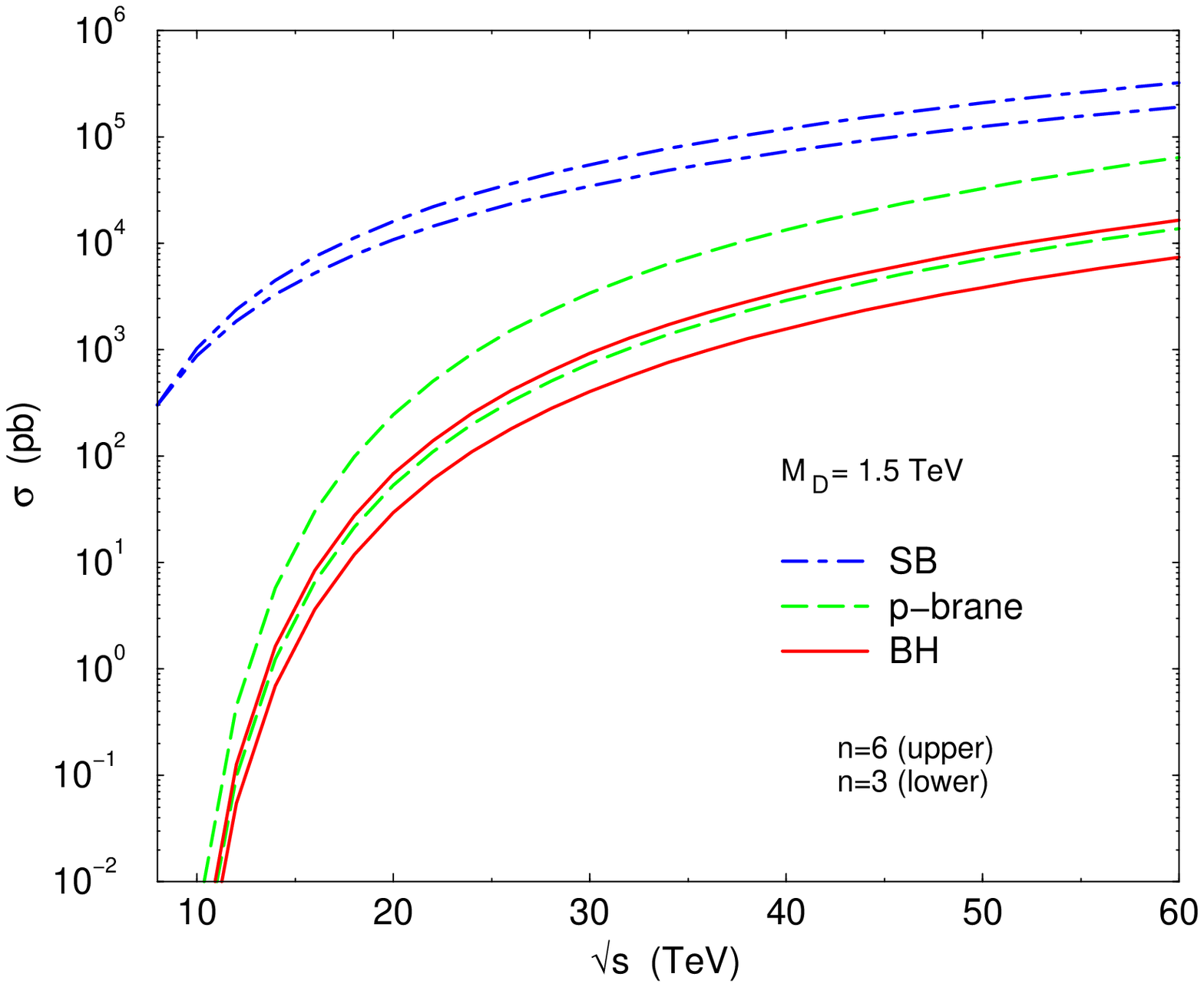}
\caption{\small
\label{total}
Total production cross section $\sigma$ including $2\to1$ and $2\to2$
processes for black hole (BH), string ball (SB), and $p$-brane ($p$B) at $pp$
collisions vs $\sqrt{s}$ for $n=3$ and $6$.
Here we have used a fundamental scale $M_D=1.5$ TeV.  The minimum mass
on the BH and $p$-brane is $M_{\rm BH}^{\rm min}, M_{p\rm B}^{\rm min}=5 M_D$,
while that on SB is $M_{\rm SB}^{\rm min}=2 M_s$.  $M_s = 1.0$ and $1.2$ TeV
for $n=3$ and $6$, respectively, in our scheme.
}
\end{figure}

In Fig. \ref{total}, we show the total production cross sections for
BH's, SB's, and $p$-branes, including the $2\to 1$ and $2\to 2$
subprocesses (when computing the $2\to 2$ subprocess we require a
$p_T$ cut of 500 GeV to prevent double counting).  Typically, the
$2\to 2$ subprocess contributes at a level of less than 10\%.  For the
BH, SB, and $p$-brane, we show the results for $n=3$ and $n=6$. The
results for $n=4,5$ lie in between.  Since we require $M_{\rm
BH}^{\rm min}, \, M^{\rm min}_{p\rm B}=5 M_D$, their production is
only sizable when $\sqrt{s}$ reaches about 10 TeV, unlike the SB,
which only requires $M_{\rm SB}^{\rm min}=2M_s$. The $p$-brane
cross section is about a few times larger than the BH, as we have
chosen $r=p=m=n-2$.  String ball production is, on average, two
orders of magnitude larger than that of a BH in the energy range
between 20 and 60 TeV.  Below 20 TeV (e.g., at the LHC), the SB
cross section is at least three orders of magnitude larger than the
BH.

\begin{figure}[th!]
\includegraphics[width=5in]{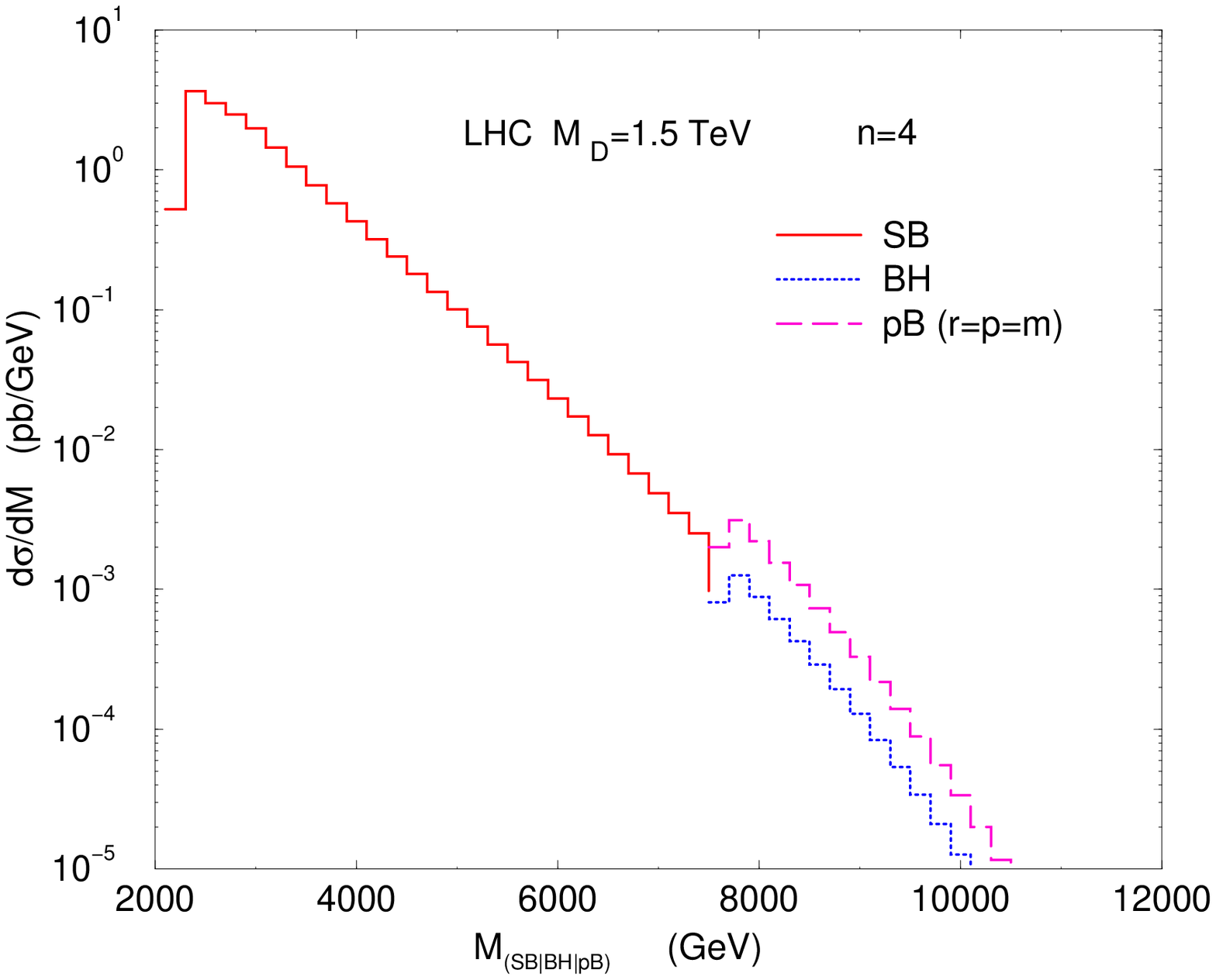}
\caption{\small
\label{dsdm}
Differential cross section $d\sigma/d M$ vs the mass $M$ of black
hole (BH), string ball (SB), or $p$-brane ($p$B) at the LHC.
Here we have used a fundamental scale $M_D=1.5$ TeV and $n=4$.  The minimum
mass on the BH and $p$-brane is $M_{\rm BH}^{\rm min}, M_{p\rm B}^{\rm min}
=5 M_D$,
while that on SB is $M_{\rm SB}^{\rm min}=2 M_s$.
$M_s = 1.1$ TeV for $n=4$.
}
\end{figure}

\begin{figure}[th!]
\includegraphics[width=5in]{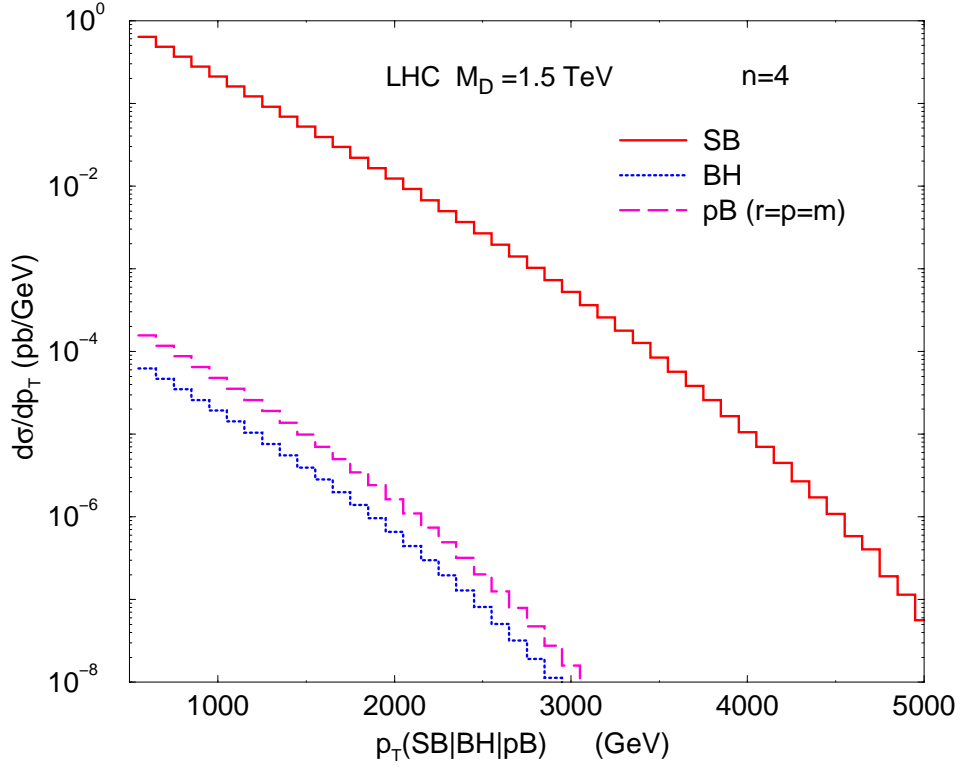}
\caption{\small
\label{dsdpt}
Differential cross section $d\sigma/d p_T$ vs the transverse momentum
 $p_T$ of black
hole (BH), string ball (SB), or $p$-brane ($p$B) at the LHC.
Here we have used a fundamental scale $M_D=1.5$ TeV and $n=4$.  The minimum
mass on the BH and $p$-brane is $M_{\rm BH}^{\rm min}, M_{p\rm B}^{\rm min}
=5 M_D$,
while that on SB is $M_{\rm SB}^{\rm min}=2 M_s$.
$M_s = 1.1$ TeV for $n=4$.
}
\end{figure}

Now we particularly look at the production rates at the LHC,
operating at $\sqrt{s}=14$ TeV with a nominal yearly luminosity of
100 fb$^{-1}$.  The differential cross sections $d\sigma/dM$,
where $M=M_{\rm BH}, M_{\rm SB}, M_{p\rm B}$, are shown in Fig.
\ref{dsdm}, where we have shown the case of $n=4$ and $M_D=1.5$
TeV.  In our scheme, $M_s \simeq 1.1$ TeV.  The minimum SB
mass starts at $2M_s \approx 2.2$ TeV, while the BH and $p$-brane
start at $5M_D=7.5$ TeV.  The SB spectrum smoothly joined to
the BH spectrum at the transition point $M_s/g_s^2=5M_D$.
Similarly, the transverse momentum spectra for their production
are shown in Fig. \ref{dsdpt}.  Even at a very high $p_T\agt 1$
TeV, the cross section is still large enough for detection.  We
show the integrated cross sections for the LHC in Table
\ref{table-lhc}, including contributions from $2\to1$ and $2\to2$
 processes (we imposed a $p_T$ cut of 500 GeV in the $2\to2$ process).

Sensitivity information can be drawn from the table.  The event rates
for BH and $p$-brane production are negligible for $M_D=2.5$ TeV and only
moderate at $M_D=2$ TeV.  At $M_D=2$ TeV, the number of BH events that
can be produced in one year running (100 fb$^{-1}$) is about $120-340$
for $n=3-7$ while the number for $p$-brane events is $210-1300$.  Therefore,
the sensitivity for a detectable signal rate for a BH and a $p$-brane is
only around 2 TeV, if not much larger than 2 TeV.
The SB event rate is much higher.  Even at $M_D=3$ TeV, the cross section
is of order of 30 pb.  In Table \ref{table-lhc}, we also show the 
$\sigma({\rm SB})$ for $M_D=4-6$ TeV.  Roughly, the sensitivity is around 
6 TeV.

\begin{figure}[th!]
\includegraphics[width=5in]{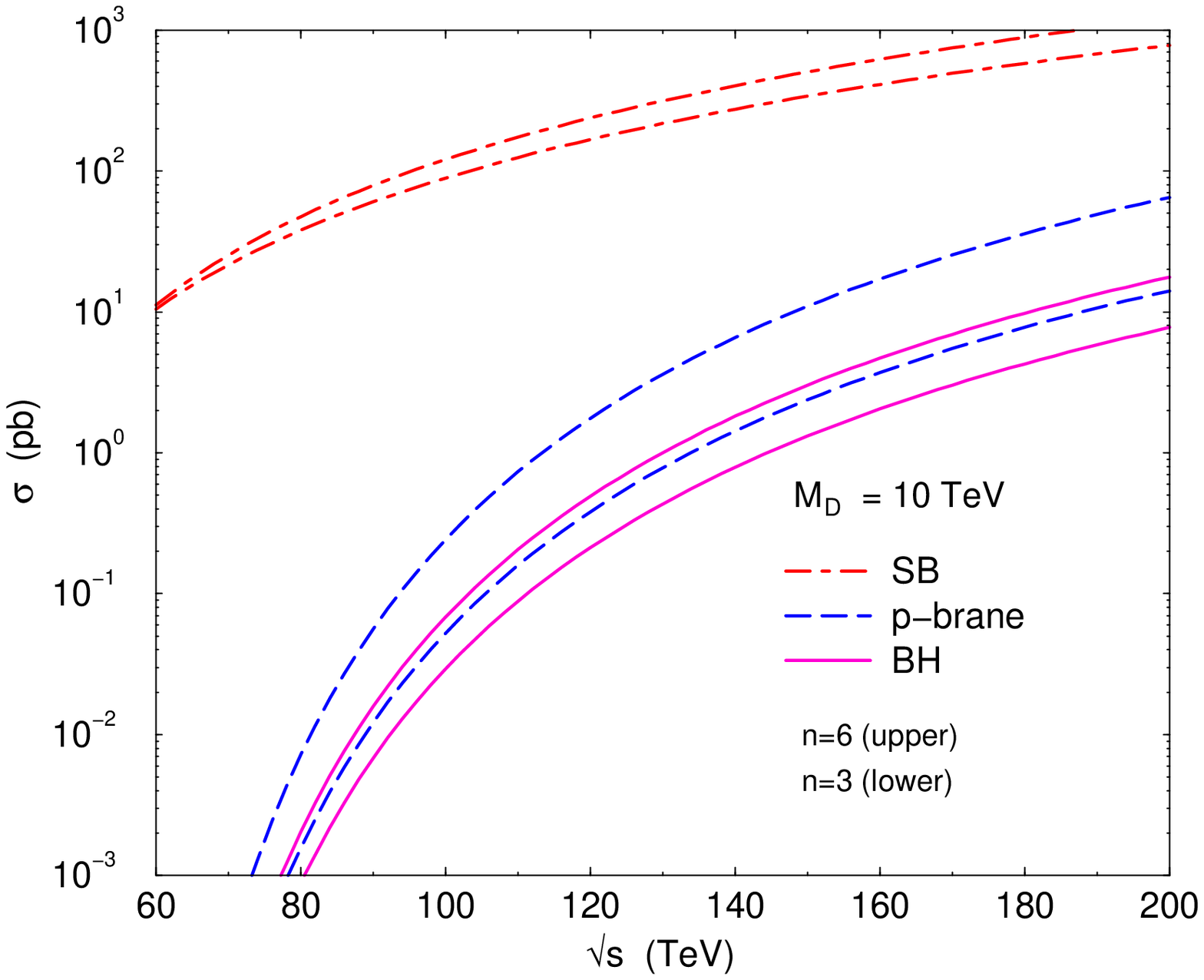}
\caption{\small
\label{total-60}
Total production cross section $\sigma$ including $2\to1$ and $2\to2$
processes for black hole (BH), string ball (SB), and $p$-brane ($p$B) at $pp$
collisions vs $\sqrt{s}$ from $60-200$ TeV for $n=3$ and $6$.
Here we have used a fundamental scale $M_D=10$ TeV.  The minimum mass
on the BH and $p$-brane is $M_{\rm BH}^{\rm min}, M_{p\rm B}^{\rm min}=5 M_D$,
while that on SB is $M_{\rm SB}^{\rm min}=2 M_s$.  
$M_s = 6.7$ and $7.9$ TeV
for $n=3$ and $6$, respectively, in our scheme.
}
\end{figure}

The VLHC (very large hadron collider) is another $pp$ accelerator
under discussions \cite{vlhc} in the Snowmass 2001 \cite{snowmass}.
The preliminary plan is to have an initial stage of about 40--60
TeV center-of-mass energy, and later an increase up to 200 TeV.  
The targeted luminosity is $(1-2)\times 10^{34}\,{\rm cm}^{-2} {\rm s}^{-1}$. 
In Fig. \ref{total-60}, we show the total production cross sections for
BH's, SB's, and $p$-branes for $\sqrt{s}=60-200$ TeV and for $n=3$ and $6$.
The integrated cross sections for $\sqrt{s}=50, \, 100, \, 150$, and $200$
TeV are shown in Table \ref{table-vlhc}.
For a fixed $M_D$, the cross section obviously increases with $\sqrt{s}$.  We 
choose to show the event rates for different values of $M_D$ such that it 
roughtly gives an idea about the sensitivity reach at each $\sqrt{s}$.  
We found that the sensitivity reaches for BH and $p$-brane production
are roughly between $6$ and $7$ TeV for $\sqrt{s}=50$ TeV,
$10$ and $13$ TeV for $\sqrt{s}=100$ TeV, 
$14$ and $18$ TeV for $\sqrt{s}=150$ TeV,
and $20$ and $25$ TeV for $\sqrt{s}=200$ TeV.  These estimates are rather crude
based on the requirement that the number of raw events is $\agt 50-100$.

\section{Decay Signatures}
\subsection{Black holes}

The main phase of the decay of a BH is via the Hawking evaporation.  The
evaporation rate is governed by its Hawking temperature, 
given by \cite{myers}
\begin{equation}
T_{\rm BH} = \frac{n+1}{4\pi R_{\rm BH} }\;,
\end{equation}
which scales inversely with some powers of $M_{\rm BH}$.  The heavier the
BH, the lower is the temperature.  Thus, the evaporation rate is slower.
The lifetime of the BH also scales inversely with the Hawking temperature
as given by
\begin{equation}
\tau \sim \frac{1}{M_D}
\left( \frac{M_{\rm BH}}{M_D} \right )^{\frac{n+3}{n+1}} \;.
\end{equation}
{}From the above equation, it is obvious  that the lifetime of a BH
becomes much longer in models of large extra dimensions than in the usual
$4D$ theory.  However, the lifetime is still so short that it will decay
once it is produced and no displaced vertex can be seen in the detector.
For another viewpoint on the BH decay, please see Ref.~\cite{casa}.

An important observation is that the wavelength $\lambda$
of the thermal spectrum corresponding to 
the Hawking temperature is larger than the size of the BH.
This implies that the BH evaporates like a point source in $s$-waves,
therefore it decays equally into brane and bulk modes, and will not
see the higher angular momentum states available in the extra
dimensions.  Since on the brane there are many more particles than
in the bulk,  the BH decays dominantly into brane modes,
i.e., the SM particles in the setup.
Furthermore, the BH evaporates ``blindly" into all degrees of freedom.
The ratio of the degrees of freedom for gauge bosons, quarks, and leptons is
$29:72:18$ (the Higgs boson is not included).  Since the $W$ and $Z$ decay
with a branching ratio of about 70\% into quarks,
 and the gluon also gives rise to hadronic activities, the final ratio
of hadronic to leptonic activities in the BH decay is about $5:1$
\cite{scott}.

 Another important property of the BH decay is the large number
of particles, in accord with the large entropy in Eq.
(\ref{entropy}), in the process of evaporation. It was shown
\cite{scott,greg} that the average multiplicity $\langle N
\rangle$ in the decay of a BH is order of $10-30$ for $M_{\rm BH}$
being a few times $M_D$ for $n=2-6$. Since we are considering
the BH that has an entropy of order 25 or more, it guarantees a
high multiplicity BH decay. The BH decays more or less
isotropically and each decay particle has an average energy of a
few hundred GeV.  Therefore, if the BH is at rest, the event is
very much like a spherical event with many particles of hundreds
of GeV pointing back to the interaction point (very much like a 
fireball).
On the other hand, if the BH is produced in
association with other SM particles (as in a $2\to k$ subprocess),
the BH decay will be a boosted spherical event on one side (a boosted
fireball), the
transverse momentum of  which is balanced by a few
particles on the other side \cite{me}.
Such spectacular events should have a negligible background.

\subsection{String balls}

Highly excited long strings emit massless quanta
with a thermal spectrum at the {\it Hagedorn} temperature.
 (The Hagedorn temperature of an excited string matches the Hawking
temperature of a BH at the corresponding point $M_{\rm BH}^{\rm min}
\equiv M_s/g_s^2$.)

At $M_{\rm SB} \alt M_s/g_s^2$, the wavelength $\lambda$ corresponding to
the thermal spectrum at the Hagedorn temperature
is larger than $R_{\rm SB}$.
This argument is very similar to that of the BH, and so
the string ball radiates like a point source and emits in $s$-waves equally
into brane and bulk modes.  With many more particles (SM particles)
on the brane than in the bulk, the SB radiates mainly into the SM
particles.

When $M_{\rm SB}$ goes below $M_s/g_s^2$, the SB has the tendency
to puff up to a {\it random-walk size} as large as the $\lambda$ of
the emissions \cite{emparan}.
  Therefore, it will see more of the higher angular momentum states
available in the extra dimensions.
Thus, it decays more into the bulk modes, but it is only temporary.
When the SB decays further, it shrinks back to the string size
 and emits as a point source again \cite{emparan}.
Most of the time the SB decays into SM particles.
On average, a SB decays into invisible quanta somewhat more often than
a BH does.

High multiplicity decay of the BH should also apply to the SB,
at least when the mass of the SB is close to the correspondence point
\cite{emparan}. Naively, we expect that 
if the SB mass decreases, the multiplicity
will decrease.  Thus, the signature of the SB is very similar to
the BH, except that it may have lower multiplicity.

\subsection{$p$-branes}
The decay of $p$-branes is not well understood, to some extent we do not
even know whether it decays or is stable.
Nevertheless, if it decays one
possibility is the decay into lower-dimensional branes, thus leading to
a cascade of branes.  Therefore, they eventually decay to a number of
 $0$-branes, i.e., BH-like objects.
This is complicated by the fact that
 when the $p$-branes decay, their masses 
might not be high enough to become BH's.
 Therefore, the final $0$-branes might be some excited string states or string
 balls. Whether the zero brane is stable or not depends on models.  Another
possibility is decay into brane and bulk particles, thus
experimentally the decay can be observed.  Or it can be a combination
of cascade into lower-dimensional branes and direct decays.
Since the size $R_{p\rm B}$ is much smaller than the size of the large
extra dimensions, we expect $p$-branes to decay mainly into brane particles.
However, the above is quite speculative.

\section{Conclusions}
In this work, we have calculated and compared the production cross
sections for black holes, string balls, and $p$-branes at hadronic
supercolliders (LHC and VLHC).  
Provided that the fundamental Planck scale is of order of 1 to a few TeV, 
large numbers of BH, SB, and $p$-brane events should be observed at the
LHC.  At the VLHC ($50-200$ TeV), the events rates are enormous.
We have also given rough estimates for the
sensitivity reaches on the fundamental Planck scale $M_D$ at various 
$\sqrt{s}$, based on the number of raw events.  The sensitivity of BH and
$p$-brane production is roughly 2 TeV at the LHC, 
$6-7$ TeV for $\sqrt{s}=50$ TeV,
$10-13$ TeV for $\sqrt{s}=100$ TeV, $14-18$ TeV for $\sqrt{s}=150$ TeV,
and $20-25$ TeV for $\sqrt{s}=200$ TeV.

Finally, we offer a few comments as follows.

\begin{enumerate}

\item  The production cross sections for BH estimated in this work
are significantly smaller than others in the literature, because we have
imposed a stringent entropy $S_{\rm BH}$ requirement on the BH.
Such a requirement is necessary to make sure the object is a BH.
Had this requirement relaxed, the cross section would have
increased substantially. For the purpose of comparing with others'
results, we also show the cross sections for smaller values of $y
\equiv M_{\rm BH}^{\rm min}/M_D,\, M_{p\rm B}^{\rm min}/M_D$ 
in Table \ref{table-y}. The cross
sections listed for $y\le 4$ should be interpreted with care,
because the smaller the ratio $M_{\rm BH}^{\rm min}/M_D$ the
stronger the string effect is and the classical description for BH
may not be valid.

\item It was pointed out in Ref. \cite{park} that a BH with an angular
momentum $J$ is likely to be formed in particle collisions when the
incoming partons are collided at an impact parameter.  In such a
case, the radius of the BH decreases and thus the naive cross-section 
formula $\hat \sigma=\pi R_{\rm BH}^2$ implies a smaller cross section
for each angular momentum $J$.  The higher the angular momentum, the larger
is the suppressions.  Nevertheless, when all $J$ (including $J=0$) 
are summed, the total cross section gives a factor of $2-3$ 
enhancement to the case of nonspinning BH.

 \item $p$-brane production is negligible if $r < p$,
because of the large volume factor suppression. But when $r=p$ (the
$p$-brane wraps entirely on the small extra dimensions of the size of the 
Planck
length), the production cross section is sizable.  Moreover,
the cross section is a few times larger than the BH production for the
case of $r=p=m$, where $m \le n-2$.

\item  The production cross section for SB's is enormous because it
does not suffer from a mass threshold as large as for the BH. The
minimum mass requirement is between $M_s$ and $M_s/g_s$.  We
typically choose $2M_s$ as the starting point for the SB.  Such a
large event rate makes the tests for string ball properties and BH
correspondence principle possible. As pointed out in Ref.
\cite{greg}, since only a very small fraction of the decay
products of a BH has missing energies, the mass of the BH can
be determined.  Moreover, the energy spectrum of the decay
products can be measured and fitted to the black-body radiation
temperature.  Thus, the Hawking radiation relationship between the
mass and temperature of a BH can be tested.  Here, similar to the
BH, both the mass and the temperature of the SB can be determined
by measuring the spectrum of the decay products.  Thus, the
relationship between the mass of the SB and the Hagedorn
temperature can be tested.

\item We have emphasized the importance and the advantages of
using the $2\to 2$ subprocess for production of BH's, SB's, and
$p$-branes, which allows a substantial transverse momentum kick to the object,
and at the same time produces an energetic high $p_T$ parton, which
provides a critical tag to the event.

 \item At the LHC and VLHC,
multiparton collisions and overlapping events may be likely to
happen. A careful discrimination is therefore necessary,
especially in the case in which the BH is produced at rest or is moving
along the beam-pipe (i.e., in $2\to 1$ subprocess).  The $2\to 2$
subprocess affords an easier signature experimentally. The high
$p_T$ parton emerging as a jet, a lepton, jets, or leptons
provides an easy tag.

\item In this study, we do not consider the difference in the
decays of BH, SB, and $p$-brane.  If we could distinguish the decay
signatures of the BH and SB, we might be able to test the BH
correspondence principle at the transition point.  We can also
test the decays of $p$-branes in more detail.

\end{enumerate}

There just appears a short review article \cite{tu} on BH production at 
hadronic colliders and by ultrahigh energy cosmic neutrinos.

\section*{Acknowledgments}
I would like to thank Jeonghyeon Song, Kang Young Lee, Yong Yeon Keum,
and Eung Jin Chun for an invitation to
the Korea Institute for Advanced Study, where part of this work was done,
 and for their hospitality.
Also thanks to Seong Chan Park for an interesting discussion.
This research was supported in part by the National Center
for Theoretical Science under a grant from the National Science
Council of Taiwan R.O.C.


\newpage

\newpage

\newpage

\newpage

\newpage

\newpage

\newpage
\begin{table}[th!]
\caption{\small \label{pbrane}
The ratio $R \equiv
\hat \sigma (M_{p\rm B}=M)/\hat \sigma (M_{\rm BH}=M)$ of Eq. (\ref{R})
for various $n$ and $p$ with $n-m\ge 2$.  We have used $M_D=1.5$ TeV
and $M_{\rm BH}=M_{p \rm B}=5 M_D$. We have assumed that the $p$-brane
wraps entirely on small extra dimensions, i.e., $r=p$.  In order to obtain
the largest ratio $R$ we have chosen $p=m$.
}
\medskip
\begin{ruledtabular}
\begin{tabular}{ccccccc}
       & $p=0$ & $p=1$ & $p=2$ & $p=3$ & $p=4$ & $p=5$  \\
\hline
$n=2$ & 1 &  &&&&  \\
$n=3$ & 1 &  1.77 &&& &\\
$n=4$ & 1 &  1.41 & 2.46 && &\\
$n=5$ & 1 &  1.25 & 1.72 & 3.02 & &\\
$n=6$ & 1 &  1.17 & 1.42 & 1.94 &3.46 & \\
$n=7$ & 1 &  1.12 & 1.27 & 1.54 & 2.10 & 3.78\\
\end{tabular}
\end{ruledtabular}
\end{table}

\begin{table}[th!]
\caption{\small \label{table-lhc}
Total cross sections in pb for the production of BH, SB, and $p$-brane,
for various values of $n$ and $M_D$ at the LHC.
The minimum mass
on the BH and $p$-brane is $M_{\rm BH}^{\rm min}, M_{p\rm B}^{\rm min}=5 M_D$,
while that on SB is $M_{\rm SB}^{\rm min}=2 M_s$.
}
\medskip
\begin{ruledtabular}
\begin{tabular}{clll}
                &  $n=3$ &  $n=5$ &  $n=7$ \\
\hline
            \multicolumn{4}{c}{\underline{BH}} \\
\underline{$M_D$ (TeV)}     &        &        &   \\
  1.5     &  $0.70$              & $1.3$ & $1.9$            \\
  2.0     &  $1.2\times 10^{-3}$ & $2.2\times10^{-3}$ & $3.4\times10^{-3}$\\
  2.5     &  $1.3\times 10^{-8}$ & $2.4\times10^{-8}$ & $3.6\times10^{-8}$\\
  \multicolumn{4}{c}{\underline{SB}} \\
\underline{$M_D$ (TeV)}     &        &        &         \\
  1.5     &  $3300$  & $4100$ & $4900$            \\
  2.0     &  $590$   & $670 $ & $760 $  \\
  2.5     &  $130$   & $130 $ & $140$  \\
  3.0     &  $33$    & $29  $ & $28$  \\
  4.0     &  $2.4$    & $1.5  $ & $1.1$  \\
  5.0     &  $0.16$    & $0.060  $ & $0.033$  \\
  6.0     &  $0.0091$    & $0.0015  $ & $0.00044$  \\
  \multicolumn{4}{c}{\underline{$p$-brane}} \\
\underline{$M_D$  (TeV)}     &        &        &         \\
  1.5     &  $1.2$  & $4.0$ & $7.6$            \\
  2.0     &  $2.1\times10^{-3}$   & $6.9\times10^{-3}$ & $0.013$  \\
  2.5     &  $2.3\times10^{-8}$   & $7.3\times10^{-8}$ & $1.4\times10^{-7}$
\end{tabular}
\end{ruledtabular}
\end{table}

\begin{table}[th!]
\caption{\small
Total cross sections in pb for the production of BH, SB, and $p$-brane,
for various values of $n$ and $M_D$ in $pp$ collisions with $\sqrt{s}=
50,100,150,200$ TeV.
The minimum mass
on the BH and $p$-brane is $M_{\rm BH}^{\rm min}, M_{p\rm B}^{\rm min}=5 M_D$,
while that on SB is $M_{\rm SB}^{\rm min}=2 M_s$.
\label{table-vlhc}} 
\begin{ruledtabular}
\begin{tabular}{clll}
                &  $n=3$ &  $n=5$ &  $n=7$ \\
\hline
            \multicolumn{4}{c}{\underline{$\sqrt{s}=50$ TeV}} \\
\underline{$M_D$ (TeV)}   & \multicolumn{3}{c}{\underline{BH}}  \\    
  5.0     &  $0.13$              & $0.24$ & $0.36$            \\
  6.0     &  $5.6\times 10^{-3}$ & $0.010$ & $0.016$\\
  7.0     &  $1.3\times 10^{-4}$ & $2.5\times10^{-4}$ & $3.7\times10^{-4}$\\
\underline{$M_D$ (TeV)}     &     \multicolumn{3}{c}{\underline{SB}} \\
  5.0     &  $370$  & $460$   & $550$            \\
  6.0     &  $130$   & $150 $ & $180 $  \\
  7.0     &  $49 $   & $55 $ & $62$  \\
\underline{$M_D$  (TeV)}    &      \multicolumn{3}{c}{\underline{$p$-brane}} \\
  5.0     &  $0.23$  & $0.73$ & $1.4$            \\
  6.0     &  $0.010$   & $0.032$ & $0.061$  \\
  7.0     &  $2.4\times10^{-4}$   & $7.6\times10^{-4}$ & $0.0014$ \\
\hline
            \multicolumn{4}{c}{\underline{$\sqrt{s}=100$ TeV}} \\
\underline{$M_D$ (TeV)}   & \multicolumn{3}{c}{\underline{BH}}  \\    
  8      &  $0.49$  & $0.91$ & $1.4$            \\
  10     &  $0.029$ & $0.055$ & $0.082$\\
  13     &  $2.2\times 10^{-4}$ & $4.2\times10^{-4}$ & $6.3\times10^{-4}$\\
\underline{$M_D$ (TeV)}     &     \multicolumn{3}{c}{\underline{SB}} \\
  8     &  $300$  & $390$   & $480$            \\
  10     &  $89$   & $110 $ & $130 $  \\
  13     &  $19 $   & $21 $ & $25$  \\
\underline{$M_D$  (TeV)}    &      \multicolumn{3}{c}{\underline{$p$-brane}} \\
  8     &  $0.89$  & $2.9$ & $5.4$            \\
  10     &  $0.053$   & $0.17$ & $0.32$  \\
  13     &  $4.0\times10^{-4}$   & $0.0013$ & $0.0024$ \\
\end{tabular}
\end{ruledtabular}
\end{table}

\begin{table}[th!]
Continue ...
\medskip
\begin{ruledtabular}
\begin{tabular}{clll}
            \multicolumn{4}{c}{\underline{$\sqrt{s}=150$ TeV}} \\
\underline{$M_D$ (TeV)}  & \multicolumn{3}{c}{\underline{BH}}  \\    
  10      &  $1.3$  & $2.4$ & $3.6$            \\
  14     &  $0.033$ & $0.061$ & $0.092$\\
  18     &  $5.6\times 10^{-4}$ & $0.0011$ & $0.0016$\\
\underline{$M_D$ (TeV)}     &     \multicolumn{3}{c}{\underline{SB}} \\
  10     &  $340$  & $450$   & $560$            \\
  14     &  $57$   & $71 $ & $ 86 $  \\
  18     &  $13 $   & $16 $ & $18$  \\
\underline{$M_D$  (TeV)}    &      \multicolumn{3}{c}{\underline{$p$-brane}} \\
  10     &  $2.4$  &  $7.7$ & $14$            \\
  14     &  $0.059$   & $0.19$ & $0.36$  \\
  18     &  $0.0010$   & $0.0032$ & $0.0061$ \\
\hline
            \multicolumn{4}{c}{\underline{$\sqrt{s}=200$ TeV}} \\
\underline{$M_D$ (TeV)}  & \multicolumn{3}{c}{\underline{BH}}  \\    
  10      &  $7.7$  & $14$ & $21$            \\
  15     &  $0.23$ & $0.43$ & $0.64$\\
  20     &  $0.0070$ & $0.013$ & $0.020$\\
  25     &  $1.4\times 10^{-4}$ & $2.5\times10^{-4}$ & $3.8\times10^{-4}$\\
\underline{$M_D$ (TeV)}     &     \multicolumn{3}{c}{\underline{SB}} \\
  10     &  $780$  & $1100$   & $1400$            \\
  15     &  $100$   & $130 $ & $ 160 $  \\
  20     &  $21 $   & $26 $ & $31$  \\
  25     &  $5.7 $   & $6.6 $ & $7.6$  \\
\underline{$M_D$  (TeV)}    &      \multicolumn{3}{c}{\underline{$p$-brane}} \\
  10     &  $14$  &  $46$ & $86$            \\
  15     &  $0.42$   & $1.3$ & $2.5$  \\
  20     &  $0.013$   & $0.040$ & $0.076$ \\
  25     &  $2.4\times10^{-4}$   & $7.8\times10^{-4}$ & $0.0015$ 
\end{tabular}
\end{ruledtabular}
\end{table}

\begin{table*}[th!]
\caption{\small \label{table-y}
Total cross sections in pb for BH and $p$-brane production 
at the LHC for various values of $y \equiv M_{\rm BH}^{\rm min}/M_D$,
$M_{p\rm B}^{\rm min}/M_D$.
}
\medskip
\begin{ruledtabular}
\begin{tabular}{c|lll|lll}
        & \multicolumn{3}{c|}{\underline{BH}} &
          \multicolumn{3}{c}{\underline{$p$-brane}}  \\
      &  $n=3$ &  $n=5$ &  $n=7$& $n=3$ &  $n=5$ &  $n=7$ \\
\hline
\underline{$M_D=1.5$ TeV} &       &        &       &  & & \\
  $y=1$   & $5700$ & $13000$ & $22000$ & 8100 & 26000 &49000 \\
  $y=2$   & $580$  & $1200$ & $2000$ &  910 & 2900 & 5600 \\
  $y=3$   & $75$  & $150$ & $230$ &     120 & 400 & 760 \\
  $y=4$   & $8.5$  & $16$ & $25$ &     $15$ & $47$ & $89$ \\
  $y=5$   & $0.70$  & $1.3$ & $1.9$ &  $1.2$ & $4.0$ & $7.6$ \\   
\underline{$M_D=2$ TeV} &  &  &   &  & & \\
  $y=1$   & $1200$ &  $2900$  & $4800$ & $1700$ & $5500$ & $10000$ \\  
  $y=2$   & $72$ & $150$ & $250$ & $110$ & $360$ & $690$  \\
  $y=3$   & $4.1$  & $8.4$ & $13$ & $6.8$ & $22$ & $42$ \\
  $y=4$   & $0.13$  & $0.26$ & $0.39$ & $0.23$ & $0.73$ & $1.4$ \\
  $y=5$   & $0.0012$  & $0.0022$ & $0.0034$ & $0.0021$ & $0.0069$ &$0.013$\\ 
\underline{$M_D=2.5$ TeV}   &      &  &   &  & & \\
  $y=1$   & $330$ & $780$ & $1300$ & $460$ & $1500$ & $2800$ \\
  $y=2$   & $10$ & $22$ & $36$ &  $16$ & $52$ & $98$ \\
  $y=3$   & $0.19$  & $0.39$ & $0.62$ & $0.32$ & $1.0$ & $1.9$ \\
  $y=4$   & $6.9\times10^{-4}$  & $0.0013$ & $0.0020$ & $0.0018$ & $0.0038$ &
$0.0072$ \\
  $y=5$   & $1.3\times10^{-8}$  & $2.4\times10^{-8}$ & $3.6\times10^{-8}$ &
 $2.3\times10^{-8}$ & $7.3\times10^{-8}$ & $1.4\times10^{-7}$ \\
\underline{$M_D=3$ TeV}   &      &  &   &  & & \\
  $y=1$   & $100$ & $250$ & $420$ & $140$ & $460$ & $880$ \\
  $y=2$   & $1.5$ & $3.2$ & $5.2$ & $2.3$ & $7.5$ & $14$ \\
  $y=3$   & $0.0057$  & $0.012$ & $0.018$ & $0.0093$ & $0.030$ &$0.057$ \\
  $y=4$   & $1.8\times10^{-7}$  & $3.5\times10^{-7}$ & $5.4\times10^{-7}$ &
            $3.1\times10^{-7}$  & $9.9\times10^{-7}$ & $1.9\times10^{-6}$ \\
  $y=5$   & - & - & - & - & - & -
\end{tabular}
\end{ruledtabular}
\end{table*}


\begin{thebibliography}{99}

\bibitem{arkani}
N. Arkani-Hamed, S. Dimopoulos, and G. Dvali, Phys. Lett {\bf B429},
263 (1998);
I. Antoniadis, N. Arkani-Hamed, S. Dimopoulos, and G. Dvali, Phys. Lett.
{\bf B436}, 257 (1998);
N. Arkani-Hamed, S. Dimopoulos, and G. Dvali, Phys. Rev. {\bf D59}, 086004
(1999).

\bibitem{hole}
P. Argyres, S. Dimopoulos, and J March-Russell, Phys. Lett. {\bf B441}, 96
(1998).

\bibitem{banks}
T. Banks and W. Fischler, hep-th/9906038.

\bibitem{emp}
R. Emparan, G. Horowitz, and R. Myers, Phys. Rev. Lett. {\bf 85}, 499 (2000).

\bibitem{scott}
S. Giddings and S. Thomas, Phys. Rev. {\bf D65}, 056010 (2002).

\bibitem{greg}
S. Dimopoulos and G. Landsberg,  Phys. Rev. Lett. {\bf 87}, 161602 (2001).

\bibitem{hoss}
S. Hossenfelder, S. Hofmann, M. Bleicher, and H. St\"{o}cker, hep-ph/0109085.

\bibitem{me}
K. Cheung, Phys. Rev. Lett. {\bf 88}, 221602 (2002).

\bibitem{giddings}
S.B. Giddings, hep-ph/0110127.

\bibitem{casa-lhc}
R. Casadio and B. Harms, hep-th/0110255.

\bibitem{hofmann}
S. Hofmann et al., hep-ph/0111052;

\bibitem{park}
S. Park and H. Song, hep-ph/0111069;

\bibitem{greg2}
G. Landsberg, Phys. Rev. Lett. {\bf 88}, 181801 (2002).

\bibitem{giudice}
G. Giudice, R. Rattazzi and J. Wells, Nucl. Phys. {\bf B630}, 293 (2002).

\bibitem{blei}
M. Bleicher et al., hep-ph/0112186;

\bibitem{solo}
S. Solodukhin, Phys. Lett. {\bf B533}, 153 (2002).

\bibitem{rizzo}
T. Rizzo, JHEP {\bf 0202}, 011 (2002).

\bibitem{RS-b}
L. Anchordoqui, H. Goldberg, and A. Shapere, hep-ph/0204228.

\bibitem{kay}
P. Jain, D. McKay, S. Panda, and J. Ralston, Phys. Lett. {\bf
B484}, 267 (2000);

\bibitem{feng}
J. Feng and A. Shapere, Phys. Rev. Lett. {\bf 88}, 021303 (2002);
L. Anchordoqui, J. Feng, H. Goldberg, and A. Shapere,
Phys. Rev. {\bf D65}, 124027 (2002).

\bibitem{anch}
L. Anchordoqui and H. Goldberg, Phys. Rev. {\bf D65}, 047502
(2002).

\bibitem{ratt}
 R. Emparan, M. Masip, and R. Rattazzi, Phys. Rev. {\bf D65},
  064023 (2002). 

\bibitem{ring}
A. Ringwald and H. Tu, Phys. Lett. {\bf B525}, 135 (2002).

\bibitem{uhe}
Y. Uehara, Prog. Theor. Phys. {\bf 107}, 621 (2002).

\bibitem{kowalski}
M. Kowalski, A. Ringwald, and H. Tu, Phys. Lett. {\bf B529}, 1
(2002).

\bibitem{han}
J. Alvarez-Muniz, J. Feng, F. Halzen, T. Han, and D. Hooper,
Phys. Rev. {\bf D65}, 124015 (2002).

\bibitem{volo}M. Voloshin, Phys. Lett. {\bf B518}, 137 (2001);
Phys. Lett. {\bf B524}, 376 (2002).


\bibitem{casa}
R. Casadio and B. Harms, Phys. Lett. {\bf B487}, 209 (2000); Phys.
Rev. {\bf D64}, 024016 (2001).

\bibitem{kanti}
P. Kanti and J. March-Rusell, hep-ph/0203223.

\bibitem{kanti0}
P. Kanti and K. Tamvakis, Phys. Rev. {\bf D65}, 084010 (2002).

\bibitem{kim}
H. Kim, S. Moon, and J. Yee, JHEP {\bf 0202}, 046 (2002).

\bibitem{ear}
D. Eardley and S. Giddings, gr-qc/020134.

\bibitem{jevi}
A. Jevicki and J. Thaler, hep-th/0203172.

\bibitem{hsu}
S. Hsu, hep-ph/0203154.

\bibitem{bilke}
S. Bilke, E. Lipartia, and M. Maul, hep-ph/0204040.

\bibitem{stoj}
V. Frolov and D. Stojkovic, e-Print Archive: hep-th/0206046.

\bibitem{emparan}
S. Dimopoulos and R. Emparan, Phys. Lett. {\bf B526}, 393 (2002).

\bibitem{ahn}
E.-J. Ahn, M. Cavaglia, and A. Olinto, hep-th/0201042.

\bibitem{jain}
P. Jain, S. Kar, S. Panda, and J. Ralston, hep-ph/0201232.

\bibitem{feng-p}
L. Anchordoqui, J. Feng, and H. Goldberg, Phys. Lett. {\bf B535}, 302 (2002).

\bibitem{oda}
K. Oda and N. Okada, hep-ph/0111298.

\bibitem{tevstring}
J. Friess, T. Han, and D. Hooper, hep-ph/0204112.

\bibitem{tye}
G. Shiu and S. Tye, Phys. Rev. {\bf D58}, 106007 (1998).

\bibitem{myers}
R. Myers and M. Perry, Ann. Phys. {\bf 172}, 304 (1986).

\bibitem{vlhc}
``M4 Working Group/Hadron Colliders", plenary talk given by M.
Syphers at the Snowmass 2001, available online at {\tt
http://www.vlhc.org/M4finalPlenary.pdf}.

\bibitem{snowmass}
Snowmass 2001 "The future of particle physics" meeting, Snowmass,
Summer 2001.

\bibitem{tu}
H. Tu, hep-ph/0205024.

\end{thebibliography}
\end{document}